\documentclass[aps,prl,showpacs]{revtex4}

\usepackage{graphicx}

\begin{document}

\title{Transition from the macrospin to chaotic behaviour by a spin-torque driven 
magnetization precession of a square nanoelement}
\author{D.~Berkov\footnote[1]{e-mail: db@innovent-jena.de}, N.~Gorn}
\affiliation{Innovent e.V., Pr\"ussingstr. 27B, D-07745, Jena, Germany}

\date{\today}

\begin{abstract}

We demonstrate (using full-scale micromagnetic simulations) that the spin injection driven
steady-state precession of a thin magnetic nanoelement exhibit a complicate transition 
from the quasi-macrospin to the chaotic behaviour with the increasing element size. 
For nanoelement parameters typical for those used experimentally we have found that
the macrospin approximation becomes invalid already for very small nanoelement sizes 
($\sim 30$ nm), in contrast to the previously reported results 
(Li and Zhang, Phys. Rev., {\bf B68}, 024404-1 (2003)).
 
\end{abstract}


\maketitle

\begin{figure}[tbhp]
\centering
{\includegraphics
[scale=0.6, bb=10cm 0cm 15cm 28cm]
{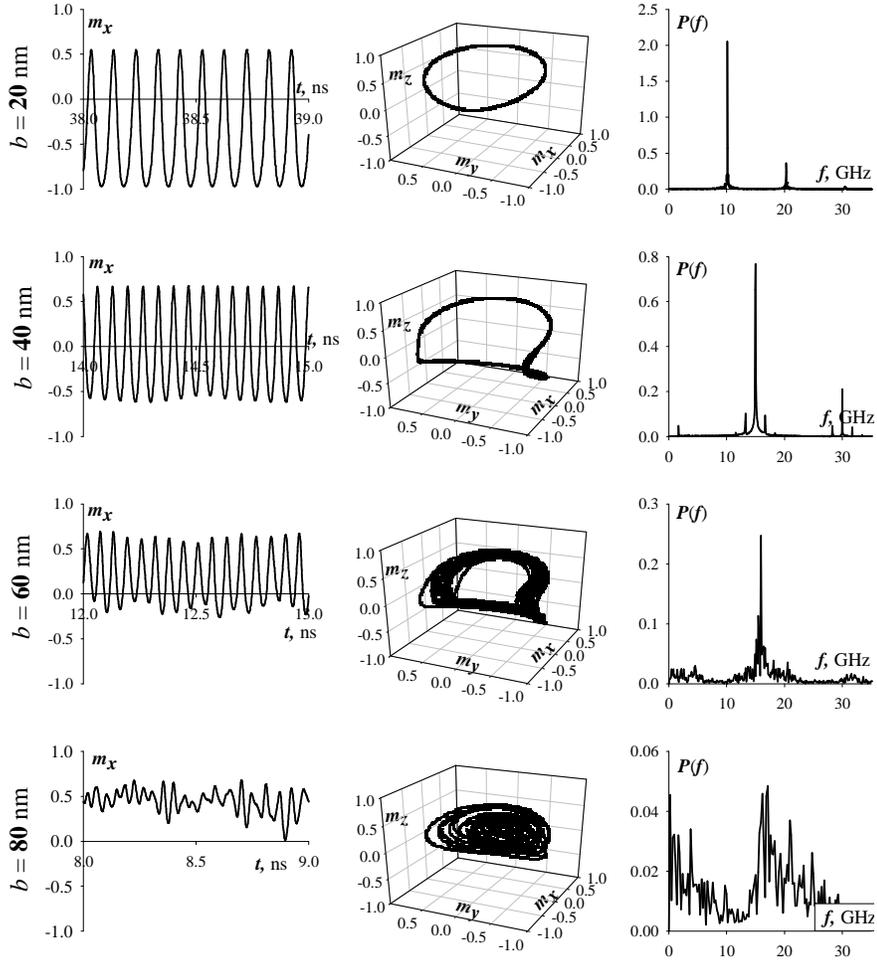}}
\caption
{Time-dependencies of the $x$-projection of the average element magnetization $m_x^{\rm av}$ 
(first column), 3D trajectories of ${\bf m}^{\rm av}$ (second column) and spectra of $m_x^{\rm av}$
(third column) for various element sizes as indicated on the left.}
\label{FigCoarseTrend} 
\end{figure}

\begin{figure}[tbhp]
\centering
{\includegraphics
[scale=0.6, bb=10cm 0cm 15cm 28cm]
{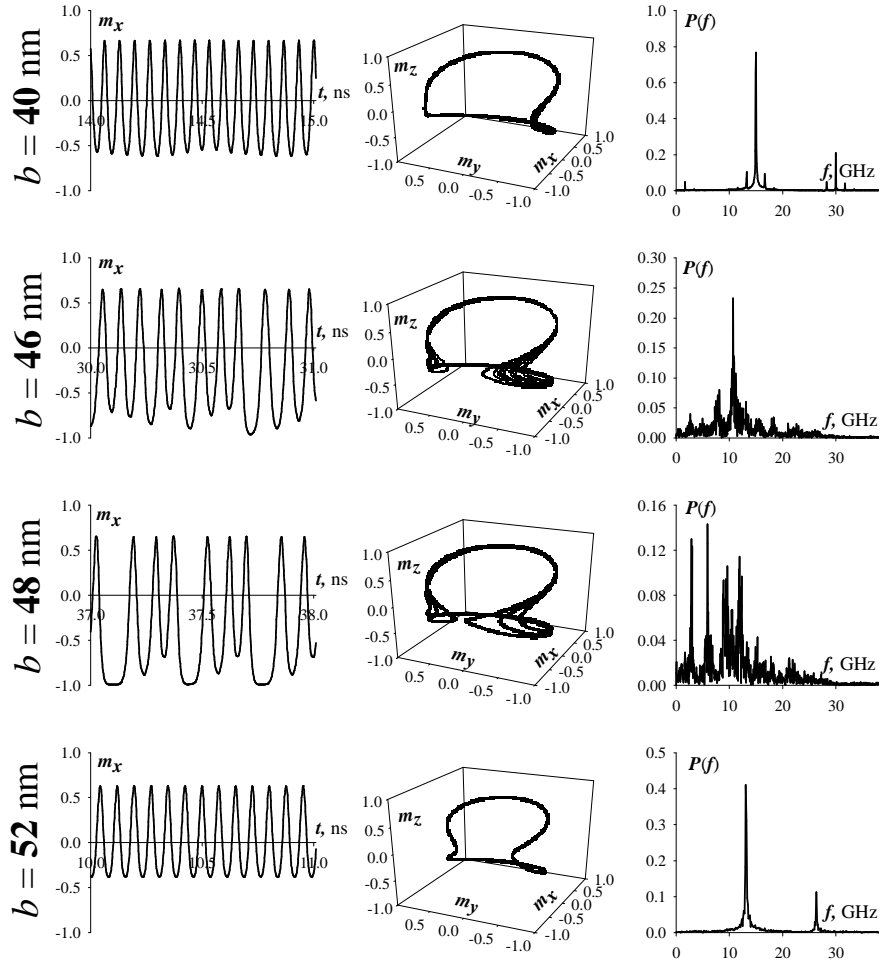}}
\caption
{The same as in Fig. \ref{FigCoarseTrend} for element sizes from $b = 40$ nm to 
$b = 52$ nm}
\label{FigFineFeature} 
\end{figure}

\begin{figure}[tbhp]
\centering
{\includegraphics
[scale=0.6, bb=2cm 0cm 15cm 20cm]
{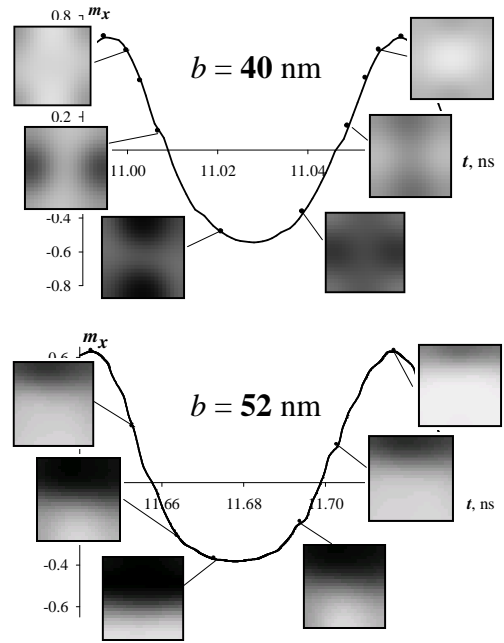}}
\caption
{Comparison of magnetization patterns during the steady-state precession for
elements with the sides $b = 40$ nm and $b = 52$ nm}
\label{CompTwoButterflies} 
\end{figure}

Magnetic excitations induced in a thin layer by a spin-polarized current injection
(first predicted theoretically \cite{SpInjPred} and soon confirmed experimentally  
\cite{SpInjDiscov}) are at present one of the most intensively studied magnetic 
phenomena due to their very interesting physical nature and highly promising potential
applications e.g., for fast switching of nanoelements \cite{SpInjSwitch} and 
design of nanosized $dc$-current driven microwave generators \cite{MWOscSpec}.
Due to the complicate remagnetization processes involved it was realized very quickly
\cite{Miltat2001} that full-scale micromagnetic simulations should be carried out to 
support corresponding experiments, because simulations in a macrospin approximation
\cite{Sun2000}, being a necessary first step in understanding some basic physics, can 
not explain many important features of experimental observations 
\cite{SpInjSwitch,MWOscSpec}.\\ 
One of the most relevant problems which could be answered by such simulations is
the determination of the critical nanoelement size for the transition from a single- 
to a multi-domain behaviour during the switching/precession process. This question
was addressed in one of the first papers where full-scale micromagnetic simulations 
of the spin-transfer induced remagnetization were carried out \cite{Li2003a}. It was 
claimed in \cite{Li2003a} that a nanoelement with magnetic parameters and thickness 
typical for real experiments remains virtually single-domain for the 
lateral size as large as $64\times 64$ nm up to the highest spin torque tested 
in the simulations (and achievable experimentally).\\ 
In this paper we present a systematic study of magnetization structures occurring
in square nanoelements of various sizes during the so called steady-state precession
\cite{Sun2000}. We have found that the transition from a quasi-macrospin precession
to a multi-domain state occurring by the increase of a nanoelement size is very complicate;
it involves several bifurcations and ends up with a fully chaotic system behaviour. 
According to our simulations (for parameters equal to those explicitly
specified in \cite{Li2003a}) the transition to a multi-domain configuration occurs
in fact already for the size $\approx 35 - 40$ nm (the difference between our results 
and those reported in \cite{Li2003a} is probably due to a much too high exchange 
constant used by Li and Zhang \cite{LiPrivComm}). \\
In our simulations of the steady-state precessional states we have used the following
system: a square-shaped monolayer element with the thickness $d = 2.5$ nm, saturation 
magnetization $M_S = 950$ G, uniaxial anisotropy with the anisotropy field
$H_K = 500$ Oe along the $0x$-axis (all these parameters are identical to those used in
\cite{Li2003a}) and the exchange constant $A = 2 \times 10^{-6}$ erg/cm. The lateral size
of a nanoelement was varied from $16 \times 16$ (where the element was clearly 
single-domain) to $120 \times 120$ nm (the transition to a chaotic state completed).
Simulations were carried out using our MicroMagus simulation package \cite{MicroMagus} 
for which we have written an additional module to include the spin injection in form of
the Slonczewski torque $\Gamma = (a_J/M_S) \cdot [{\bf M} \times [{\bf M} \times {\bf S}]]$ 
(${\bf S}$ is the spin polarization direction of the current through a layer).
All results presented here were obtained for ${\bf S}$ and the external field 
$H_{\rm ext} = 1000$ Oe both directed along the $0x$-axis and the spin current strength 
$a_J = 0.4M_S$. The Oersted field was not included to make the system as similar
as possible to that studied in \cite{Li2003a}.
The lateral mesh size mostly used was $2 \times 2$ nm. We have checked 
that all results were nearly independent on the discretization: doubling the
number of discretization cells along each side never led to any qualitative changes
of the magnetization configurations or power spectra (see below) and resulted in 
the shift of the spectral peak positions up to maximum 5 \%.\\
The coarse trend describing the system behaviour when the lateral size $b$ of the square 
element increases is relatively simple. For very small sizes (up to $b \approx 20$ nm) 
the element behaves 
itself as a macrospin (first row in Fig. \ref{FigCoarseTrend}) - the magnetization 
configuration for any time remains almost collinear. The 3D trajectory of the average 
magnetization ${\bf m}^{\rm av}$ shows a well known 'out-of-plane' precession which was 
predicted by both single-spin \cite{Sun2000} and finite-element micromagnetic simulations 
\cite{Li2003a}. The power spectrum of $m_x^{\rm av}$-oscillations in this case exhibits 
a very sharp peak at the precession frequency $f_{\rm prec}$ and much smaller peaks with rapidly 
decreasing amplitudes by $2f_{\rm prec}$, $3f_{\rm prec}$, etc. due to the slightly 
non-sinusoidal character of the precession.\\
When the element size is increased, the maximal precession angle also increases until the
'out-of-plane' precession trajectory touches the element ($0xy$) plane (this takes place 
for $b = 32 \pm 1$ nm) and the transition to the 'butterfly' trajectory occurs.
The magnetization structure exhibits clear deviations from the single-domain one:
the maximal angle between magnetization vectors at various element points can exceed
$45^o$. Although the total oscillation period {\it increases} (the limiting cycle is now much 
longer than for a simple quasi-elliptical 'out-of-plane' precession), the spectrum peak 
of $m_x^{\rm av}$-oscillations moves towards {\it higher} frequencies because one cycle includes 
now {\it two} oscillation periods for $m_x^{\rm av}$ (second row in Fig. \ref{FigCoarseTrend}). The 
weak satellites of the main spectral lines are due to a slow back-and-forth displacement 
of the ${\bf m}^{\rm av}$-trajectory in the region near the $0xy$-plane which is due to 
the fact that the energy landscape there is nearly flat.\\
Further enlargement of the element leads to the formation of well defined domains with sharp
domain walls between them. This, in turn, results in the transition to the {\it quasiperiodic} 
${\bf m}^{\rm av}$-trajectory, which now fills a bended torus (third row for $b=60$ nm 
in Fig. \ref{FigCoarseTrend}). The co-existence of several domains explains the broadening 
of the corresponding spectral line and the shift of the $m_x^{\rm av}(t)$-dependence towards
higher $m_x^{\rm av}$-values - due to the external field in the positive $0x$-direction.\\
For still larger element sizes the transition to the {\it chaotic} behaviour finally occurs (see
example for $b=80$ nm in the fourth row in Fig. \ref{FigCoarseTrend}).
The magnetization trajectory completely fills in an area near the pole $m_x=1$, $m_y=m_z=0$.
The size of the filled area decreases with the increasing element size due to the same 
aligning effect of the external field. The spectral power is gradually transferred to very 
low frequencies, because significant changes of the average $m_x^{\rm av}$-value at a time
scale much larger than the (still visible) oscillation period occur due to the chaotic
domain structure of the nanoelement.\\
Simulations for several intermediate system sizes reveal that the transition from the 
single-domain to the chaotic behaviour described above contains a very interesting 
intermediate stage (Fig. \ref{FigFineFeature}). First of all we note that although the 
qualitative behaviour of the {\it average} magnetization for the sizes, e.g., $b = 40$ nm 
and $b = 52$ nm is the same - the 3D ${\bf m}^{\rm av}$-trajectory of the 'butterfly' 
type is observed and corresponding spectra are quite similar (the first and the last rows in 
Fig. \ref{FigFineFeature}), magnetization patterns during the precession are very 
different (Fig. \ref{CompTwoButterflies}).\\
For the smaller size $b = 40$ nm the
magnetization direction inside the element remains roughly the same (maximal angle
between ${\bf m}({\bf r})$-vectors at various element points is $\sim 50^o$). Pairs of
semicircular 'quasidomains' are formed near the opposite element sides during the 
precession (Fig. \ref{CompTwoButterflies}, upper graph for $b = 40$ nm) and transitions 
between the areas with different ${\bf m}({\bf r})$-directions are very smooth.\\
In contrast, magnetization configuration for the $b = 52$ nm 
element exhibits two sharply defined domains with nearly opposite magnetization
directions (Fig. \ref{CompTwoButterflies}, lower picture for $b = 52$ nm). 
The corresponding domain wall moves up and down during the steady-state 
precession resulting in the overall change of ${\bf m}^{\rm av}$. The physical reason 
for this difference probably is that due to the larger element size it is energetically 
more favourable to form two relatively well defined and nearly homogeneously 
magnetized domains than to keep the configuration where 
the magnetization direction varies smoothly along the whole nanoelement.\\
The transition between the two 'butterfly'-regimes described above occurs in a very 
interesting way. When the size is increased above 
$b = 40$ nm, the limiting cycle broadens into a band of trajectories which width
is maximal in the already mentioned critical region near the $0xy$-plane 
(see 3D-picture for $b = 46$ nm in the middle of the second row in
Fig. \ref{FigFineFeature}); this also results in the shift and the broadening
of the main spectral line. By further (very small !) size increase up to $b = 48$ nm 
the continuous band of trajectories splits itself and all trajectories collapse
into 3 limiting subcycles (third row in Fig. \ref{FigFineFeature}). A peak at a frequency
much lower than that of the 'normal' precession corresponding to the complete motion
cycle over all these three subcycles appears in the spectrum together with its harmonics. 
And finally, the size increase up to $b = 52$ nm let all the subcycles collapse into 
a single 'butterfly'-type limiting cycle (see above).\\
The detailed discussion of these results will be given elsewhere. Here we only
mention that, according to our simulations, the transitions described above 
are shifted, as expected, towards higher lateral sizes when the exchange constant
or the element thickness is increased. Furthermore, the smoothing of the square 
corners also stabilizes to some extend the homogeneous magnetization configuration.
However, for all studied values of the exchange constants (up to $A = 4 \times 10^{-6}$ erg/cm),
all thicknesses up to $d = 5$ nm and all smoothing radii (up to the circular shape) the
element with the lateral size $b = 120$ nm was always at least in a multi-domain 
state. We also note that an inclusion of the Oersted field shifts all transitions towards
{\it smaller} sizes because this field is strongly inhomogeneous. \\
Some brief comments concerning theoretical interpretation of our results and their 
relation to experimental observations are in order. First we note that a theoretical 
analysis of our simulation data obviously requires the application of the non-linear
dynamics formalism \cite{NonLinDynBook}. For example, qualitative changes of the 
magnetization trajectories shown in Fig. \ref{FigFineFeature} strongly resemble 
pictures of processes known as 'period multiplication' bifurcations; time-dependent 
trajectories for, e.g., $b = 48$ nm are very similar to those known from the Lorentz 
attractor studies \cite{NonLinDynBook}; and the overall behaviour of the system when 
the size is increased from the smallest $b = 20$ nm to the largest one $b = 120$ nm 
obviously represents a nice example of the transition from a regular to chaotic 
behaviour when one of the system parameters (the nanoelement 
size in our case) is varied \cite{NonLinDynBook}.\\ 
However, we point out that the corresponding rigorous analysis using the non-linear 
dynamics terms will be extremely difficult. The reason is that our system is actually 
a $2N$-dimensional one, where $N = N_x \times N_y$ is the total number of the 
discretization cells. This means that trajectories of the {\it average} magnetization 
shown in Fig. \ref{FigCoarseTrend} and \ref{FigFineFeature} are {\it not} the trajectories 
in the system phase space which are required for the analysis with the non-linear 
dynamics methods, but merely the trajectories of a certain low-dimensional (3D)
functionals defined on this phase space. In the same fashion, the analysis of
the system behaviour in terms of Ljapunov exponents would require the (very 
precise !) determination of the eigenvalues of the corresponding $2N \times 2N$
matrices which is a tedious computational task, leaving apart a rather non-trivial
interpretation problem of the resulting sets of $2N$ eigenvalues. For these reasons
we believe that such studies should be carried out only if one expects from them
new deep insights into the physics of the process beyond those which can be obtained
by studying the standard physical characteristics like discussed above.\\
Results like those shown here may be in principle compared {\it directly} with 
the experimental data - especially since high-quality measurements of the spin-transfer 
induced microwave oscillation spectra for various systems are available \cite{MWOscSpec}.
However, our simulations demonstrate that for the nanoelement sizes commonly used
in such experiments, the nature of the precessional state (which is necessary to 
ensure the very existence of the non-decaying m/w oscillations) is very sensitive
to the element size. Furthermore (corresponding results will be reported elsewhere)
the shape of the element and its polycrystalline structure (for materials with 
significant magnetocrystalline anisotropy) also play an important role. Apart from
the evident consequence that simulations of a spin-torque induced precession of 
nanoelements with experimentally interesting sizes in a macrospin approximation are 
not very helpful, our results show that a {\it precise} determination of all 
mentioned above physical parameters of the experimentally studied system  
is required to enable a meaningful comparison with micromagnetic simulations. 
The latter step is, in turn, necessary for a real understanding of physical processes 
underlying the magnetization dynamics in the presence of a spin-polarized current.\\

\newpage

\end{document}